\documentclass[notitlepage,nofootinbib,preprintnumbers,amssymb,superscriptaddress,twocolumn]{revtex4-2}
\usepackage{amsfonts,amssymb,amsmath,mathtools,graphicx,color,bm,orcidlink}
\definecolor{ultramarine}{rgb}{0.07, 0.04, 0.56}
\definecolor{cadmiumgreen}{rgb}{0.0, 0.42, 0.24}
\definecolor{indigo(dye)}{rgb}{0.0, 0.25, 0.42}
\usepackage{hyperref}
\hypersetup{
colorlinks=true,
citecolor=ultramarine,
linkcolor=cadmiumgreen,
urlcolor=indigo(dye),
}

\let\originalleft\left
\let\originalright\right
\renewcommand{\left}{\mathopen{}\mathclose\bgroup\originalleft}
\renewcommand{\right}{\aftergroup\egroup\originalright}
\usepackage{autobreak}
\usepackage{enumitem}

\allowdisplaybreaks

\begin{document}

\preprint{RESCEU-13/26}

\title{Relativistic frequency shifts in gravitational waves from axion clouds}

\author{Takuya Takahashi\,\orcidlink{0000-0002-4894-6108}}
\email{takuya.takahashi@resceu.s.u-tokyo.ac.jp}
\affiliation{Research Center for the Early Universe (RESCEU), Graduate School of Science, The University of Tokyo, Tokyo 113-0033, Japan}

\begin{abstract}
Superradiant instability of ultralight bosons can produce clouds around rotating black holes, whose continuous gravitational wave (GW) emission is a promising observational target. Precise predictions of the signal frequency and its evolution are essential for detecting such continuous GWs. For axions, self-interactions can populate multiple superradiant modes via nonlinear couplings, and GW emission can occur through various channels. To calculate the frequency shifts of GWs emitted through these channels, we employ relativistic perturbation theory based on a bilinear form. We apply this framework to self-interaction effects for the first time, and also revisit the treatment of the self-gravity contribution. Our results provide a simple and unified framework for calculating frequency shifts, including cases in which multiple modes are excited, and are relevant for next-generation GW observations.
\end{abstract}

\maketitle

\section{Introduction}\label{sec:intro}
Gravitational wave (GW) observations have created new opportunities to search for fundamental particles that are difficult to access in laboratory experiments.
Ultralight bosons are among the best-motivated possibilities in this direction.
Axions are particularly interesting since they arise as a solution to the strong CP problem~\cite{Peccei:1977hh,Weinberg:1977ma,Wilczek:1977pj,Kim:2008hd} and may account for dark matter~\cite{Dine:1982ah,Preskill:1982cy,Abbott:1982af,Hui:2016ltb}.
Moreover, string theory generically predicts many axion-like fields in the so-called axiverse scenario~\cite{Arvanitaki:2009fg,Svrcek:2006yi,Mehta:2021pwf}.

Rotating black holes (BHs) can trigger the superradiant instability of ultralight bosons~\cite{Press:1972zz,Zouros:1979iw,Detweiler:1980uk,Dolan:2007mj,Brito:2020oca}.
In this process, a bosonic field gravitationally bound to the BH repeatedly extracts energy and angular momentum, and a macroscopic condensate forms around the BH.
The instability is most efficient when the Compton wavelength is comparable to the BH radius, corresponding to boson masses in the range $10^{-20}$-$10^{-10}$ eV for astrophysical BHs.
The presence of such a cloud can produce a variety of observational signatures~\cite{Arvanitaki:2010sy}, including BH spin-down~\cite{Brito:2014wla,Stott:2020gjj}, impacts on GWs from binaries~\cite{Baumann:2018vus,Baumann:2019ztm,Baryakhtar:2022hbu,Dyson:2025dlj}, effects on the BH shadow~\cite{Roy:2021uye,Chen:2019fsq,Chen:2022oad,Chen:2022nbb}, and a stochastic GW background~\cite{Brito:2017wnc,Tsukada:2018mbp,Gavilan-Martin:2026zzw}.

As present and future GW detectors continue to improve in sensitivity, GWs emitted from boson clouds are becoming an increasingly important target~\cite{LIGOScientific:2021rnv,KAGRA:2022osp,LIGOScientific:2025csr,Yoshino:2013ofa,Yoshino:2014wwa,Brito:2017zvb,Palomba:2019vxe,Zhu:2020tht}.
GWs from boson clouds can be broadly divided into two classes: continuous signals emitted spontaneously over long timescales, and burst-like signals associated with the collapse of the cloud, often referred to as a bosenova~\cite{Arvanitaki:2010sy,Yoshino:2012kn,Yoshino:2015nsa}.
The latter is expected to occur only in limited situations, such as in binary systems~\cite{Takahashi:2024fyq}, and in this work we focus on the former.
Continuous GWs can further be separated into two channels~\cite{Collaviti:2024mvh,Omiya:2024xlz}: pair annihilation, in which two bosons convert into a graviton, and level transitions, where GWs are emitted through transitions to another superradiant mode.

Most existing observational constraints have focused on the pair annihilation channel.
In the case of axions, however, self-interactions are generically present and can play an important role in the evolution of the cloud~\cite{Gruzinov:2016hcq,Baryakhtar:2020gao,Omiya:2022gwu,Witte:2024drg}.
In particular, they induce dissipation and energy transfer between different modes through mode couplings, which can realize quasi-stationary states in which multiple modes are simultaneously excited.
In such situations, GWs from level transitions can become significant and can even exceed those from pair annihilation.
Although the corresponding frequencies are lower, typically in the range ${\cal O}(0.1)$-${\cal O}(10)$ Hz, they could be probed by next-generation ground-based detectors such as the Einstein Telescope~\cite{Punturo:2010zz} and the Cosmic Explorer~\cite{Reitze:2019iox}, as well as space-based detectors such as DECIGO~\cite{Kawamura:2011zz}.

In practical searches for continuous GWs, accurate modeling of the signal frequency and its evolution is crucial.
This requires calculating frequency shifts induced by various perturbations.
In particular, since the shifts vary with the cloud mass, the effects of self-gravity and self-interaction become important.
Calculations of these shifts in the non-relativistic approximation were presented in Ref.~\cite{Baryakhtar:2020gao}.
(The shift due to self-angular momentum is given in Ref.~\cite{Takahashi:2021yhy}.)
For self-gravity, the frequency evolution has also been studied in numerical relativity simulations~\cite{Siemonsen:2022yyf,May:2024npn}.
For self-interaction, corrections up to second order in the cloud mass were derived using the renormalization group method~\cite{Omiya:2020vji}, and a fully nonlinear adiabatic approach tracking the joint evolution of the cloud mass and frequency has also been developed~\cite{Omiya:2022mwv}.

However, the methods described above have mainly been applied to situations in which a single mode is dominant.
In order to analyze GWs from more general configurations, such as axion clouds with multiple excited modes generated by self-interactions, a more efficient framework is needed.
In this work, we adopt the relativistic perturbation theory based on the bilinear form introduced in Ref.~\cite{Green:2022htq} and later applied to boson clouds in Ref.~\cite{Cannizzaro:2023jle} (see also Refs.~\cite{Cannizzaro:2025vpb,Minucci:2026dgo} for recent applications).
We apply this framework to frequency shifts induced by self-interactions for the first time.
We also revisit the self-gravity contribution, treating the gravitational perturbation in the Newtonian approximation as in Ref.~\cite{Cannizzaro:2023jle}, but with several improvements: the inclusion of spatial metric perturbations, a relativistic treatment of the source, and higher multipole moments.
In addition, we explicitly compute the GW frequencies for several processes of observational interest.

This paper is organized as follows.
In Sec.~\ref{sec:review}, we briefly review axion clouds formed via superradiance, their self-interaction effects, and the associated GW emissions.
In Sec.~\ref{sec:perturbation}, we present the formulation of relativistic perturbation theory and derive the frequency shift formula.
In Sec.~\ref{sec:numerical}, we provide the results of numerical computations.
Finally, we give our conclusions in Sec.~\ref{sec:conc}.
Throughout this paper, we work with the metric signature $(-,+,+,+)$ and use units with $c=G=\hbar=1$, unless otherwise stated.

\section{Review of axion clouds}\label{sec:review}
In this section, we summarize the essentials of axion clouds around rotating BHs and their GW emission.
This is intended to provide context for the results of this study and to introduce several useful definitions.

\subsection{Superradiance}
First, we consider a massive real scalar field on a Kerr spacetime.
The Kerr metric is given by
\begin{align}
    &g_{\mu\nu}dx^\mu dx^\nu \notag \\
    &=-\left(1-\frac{2Mr}{\Sigma}\right)dt^2-\frac{4Mar\sin^2\theta}{\Sigma}dtd\varphi+\frac{\Sigma}{\Delta}dr^2 \notag \\
    &\phantom{=}+\Sigma d\theta^2+\left(r^2+a^2+\frac{2Ma^2r}{\Sigma}\sin^2\theta\right)\sin^2\theta d\varphi^2~,
\end{align}
where
\begin{align}
    \Delta=&r^2-2Mr+a^2~, \\
    \Sigma=&r^2+a^2\cos^2\theta~,
\end{align}
in Boyer--Lindquist coordinates.
Here, $M$ is the BH mass and $a$ is the spin parameter.
The roots of $\Delta=0$, $r_\pm=M\pm\sqrt{M^2-a^2}$, define the horizon radii, where $r_+$ and $r_-$ denote the radius of event horizon and the Cauchy horizon, respectively.

For the scalar field, the equation of motion is given by
\begin{align}\label{eq:KGeq}
    \left(\Box_g-\mu^2\right)\phi=0~,
\end{align}
where $\mu$ is the mass of the scalar field.
The solution can be decomposed into variables as~\cite{Brill:1972}
\begin{align}
    \phi=e^{-i(\omega t-m\varphi)}R_{lm\omega}(r)S_{lm\omega}(\theta)+{\rm c.c.}~,
\end{align}
where ${\rm c.c.}$ denotes the complex conjugate.
On a Kerr background, quasi-bound states satisfying the following boundary conditions are allowed:
\begin{align}
    R_{lm\omega}(r)\to
    \begin{cases}
        \left(r-r_+\right)^{-i\frac{2Mr_+}{r_+-r_-}(\omega-\Omega_H)} 
        & (r\to r_+)~, \\
        r^{-1} r^{\frac{M(\mu^2-2\omega^2)}{\sqrt{\mu^2-\omega^2}}} e^{-\sqrt{\mu^2-\omega^2}r} 
        & (r\to \infty)~,
    \end{cases}
\end{align}
where $\Omega_H=a/2Mr_+$ is the angular velocity of the horizon.
The eigenstates and the corresponding eigenfrequencies $\omega_{nlm}$ are specified by the quantum numbers $(n,l,m)$, with $n$, $l$, and $m$ denoting the principal, orbital angular momentum, and magnetic quantum numbers, respectively~\cite{Baumann:2019eav}.
Following the hydrogen atom case, we take $n\geq l+1$.

In general, the eigenfrequencies take complex values and are written as
\begin{align}
    \omega_{nlm}=\omega_{nlm,R}+i\omega_{nlm,I}~.
\end{align}
In particular, if the superradiant condition
\begin{align}\label{eq:SRcond}
    0<\omega_{nlm,R}<m\Omega_H
\end{align}
is satisfied, then $\omega_{nlm,I}>0$.
Therefore, bound states that satisfy $\omega_{nlm,R}<\mu$ grow without losing energy to infinity, which is referred to as the superradiant instability.
As a result, starting from natural initial conditions such as quantum fluctuations~\cite{Fu:2025ztk}, a scalar cloud forms around the BH.

The energy--momentum tensor of the scalar field is given by
\begin{align}\label{eq:energy_momentum_tensor}
    T_{\mu\nu}=\nabla_\mu\phi\nabla_\nu\phi+g_{\mu\nu}\left(-\frac{1}{2}\nabla^\rho\phi\nabla_\rho\phi-\frac{\mu^2}{2}\phi^2\right)~,
\end{align}
from which the cloud mass can be defined as
\begin{align}
    M_c&=-\int_{\Sigma_t} d^3x\sqrt{-g}T^t_{\ t} \notag \\
    &=\int_{r_+}^{\infty}dr\int_{S^2}d^2\Omega \notag \\
    &\phantom{=}\times\left[\frac{1}{2\Delta}\left((r^2+a^2)^2-\Delta a^2\sin^2\theta\right)(\partial_t\phi)^2\right. \notag \\
    &\phantom{=\int}+\frac{\Delta}{2}(\partial_r\phi)^2+\frac{1}{2}(\partial_\theta\phi)^2+\frac{1}{2}\left(\frac{1}{\sin^2\theta}-\frac{a^2}{\Delta}\right)(\partial_\varphi\phi)^2 \notag \\
    &\phantom{=\int}\left.+(r^2+a^2\cos^2\theta)\frac{\mu^2}{2}\phi^2\right]~.
\end{align}
where $d^2\Omega=\sin\theta d\theta d\varphi$.
In the absence of any effects that suppress the growth, the cloud continues to extract energy and angular momentum until the BH spins down and the superradiant condition~\eqref{eq:SRcond} is saturated.
At that point, the cloud mass can reach up to $\sim10\%$ of the BH mass~\cite{Brito:2017zvb}.
Afterward, it emits GWs and slowly dissipates.
However, once scalar self-interactions are taken into account, the evolution of the system and the associated GW emission become more complex.

\subsection{Self-interaction}
As the field amplitude grows, nonlinear self-interactions play an important role in the evolution of the system.
The potential of the scalar field appears in the action as
\begin{align}\label{eq:action_SI}
    S=\int d^4x\sqrt{-g}\left(-\frac{1}{2}g^{\mu\nu}\nabla_\mu\phi\nabla_\nu\phi-V(\phi)\right)~,
\end{align}
and for axion it typically takes the form
\begin{align}
    V(\phi)&=\mu^2 F_a^2\left(1-\cos\frac{\phi}{F_a}\right) \notag \\
    &\sim \frac{1}{2}\mu^2\phi^2-\frac{1}{4!}\mu^2F_a^2\left(\frac{\phi}{F_a}\right)^4~,
\end{align}
where $F_a$ is the decay constant.
Since the $\phi^4$ term gives the dominant contribution to the evolution of the system, as shown in the literature~\cite{Omiya:2022mwv}, we truncate at this order.
Then, the equation of motion becomes
\begin{align}
    \left(\Box_g-\mu^2\right)\phi=-\frac{\mu^2}{3!F_a^2}\phi^3~.
\end{align}

Here we denote the quasi-bound eigenmodes of the homogeneous equation by
\begin{align}
    &\phi_i=\frac{1}{\sqrt{2}}\left(\Phi_i+\Phi_i^\ast\right)~, \\
    &\Phi_i=\sqrt{M_{c,i}}\ e^{-i(\omega_i t-m_i\varphi)}R_i(r)S_i(\theta)~, \label{eq:mode}
\end{align}
where $i$ labels the eigenmodes corresponding to $(n,l,m)$ and ${}^\ast$ denotes complex conjugation.
Nonlinear terms give rise to mode coupling, leading to dissipation of the cloud and energy transfer among modes.
The main dissipative processes are absorption into the BH via non-superradiant modes and radiation to infinity via unbound states, both of which are excited by the self-interaction.
Writing $\phi=\phi_i+\phi_j+\phi_k+\cdots$, the time evolution of the cloud mass for each mode is given by
\begin{align}
    \frac{dM_{c,i}}{dt}=&2\omega_{i,I}M_{c,i} \notag \\
    &+\sum_{j,k}\left[-\frac{(1+\delta_{ij})\omega_{i,R}}{\omega_{i,R}+\omega_{j,R}-\omega_{k,R}}(F_{ijk^\ast}^{\cal I}+F_{ijk^\ast}^{\cal H})\right. \notag \\
    &\phantom{+\sum_{j,k}\left[\right.}\left.+\frac{\omega_{i,R}}{\omega_{j,R}+\omega_{k,R}-\omega_{i,R}}(F_{jki^\ast}^{\cal I}+F_{jki^\ast}^{\cal H})\right] \notag \\
    &\phantom{+\sum_{j,k}\left[\right.}\times M_{c,i}M_{c,j}M_{c,k}~,
\end{align}
where $F_{ijk^\ast}^{\cal I}$ and $F_{ijk^\ast}^{\cal H}$ denote the coefficients associated with the axion energy flux to infinity and into the horizon, respectively.
The first term in the bracket represents the energy outflow to the $k$-mode, while the second term represents the energy inflow to the $i$-mode.
By solving simultaneously for the BH mass and angular momentum, including the effects of the flux into and out of the BH, one can obtain the adiabatic evolution of the system.
For details, see Ref.~\cite{Omiya:2024xlz}.
An important consequence is that, once self-interaction is taken into account, the cloud can develop a configuration in which multiple modes coexist, with a complicated dependence on the parameters.

\subsection{Gravitational wave}
Since the axion cloud is oscillating, it emits GWs.
The decay timescale through GW emission is much longer than the superradiant growth timescale and can therefore be neglected in the evolution, although it still can produce observable signals.
More specifically, the time--dependent oscillatory components of the axion energy--momentum tensor $T_{\mu\nu}$ in Eq.~\eqref{eq:energy_momentum_tensor} serve as a source of GWs.

This GW emission has two channels.
One is the ``pair annihilation" channel, sourced by the $\Phi_i\Phi_j$ components of $T_{\mu\nu}$, with frequency $\omega=\omega_{i,R}+\omega_{j,R}$.
The other is the ``level transition" channel, sourced by the $\Phi_i\Phi_j^\ast$ components of $T_{\mu\nu}$, with frequency $\omega=\omega_{i,R}-\omega_{j,R}$.
The amplitude of each GW signal is proportional to $\sqrt{M_{c,i}M_{c,j}}$, and its explicit value is obtained by solving the Teukolsky equation with the corresponding source term~\cite{Omiya:2024xlz}.

Here, we provide estimates of representative GW frequencies from each channel.
For the pair annihilation channel, since $\omega_{i,R}\sim\mu$, the frequency is given by
\begin{align}
    f_{\rm ann}\sim 646\ {\rm Hz}\ \left(\frac{10M_\odot}{M}\right)\left(\frac{M\mu}{0.1}\right)~.
\end{align}
For the level transition channel, analyses including the evolution of the cloud have shown that the signal from the $(5,4,4)\to(3,2,2)$ transition is expected to be dominant.
The corresponding frequency is given by
\begin{align}
    f_{\rm trans}\sim 0.11\ {\rm Hz}\ \left(\frac{10M_\odot}{M}\right)\left(\frac{M\mu}{0.1}\right)^3~.
\end{align}
These estimates are based on the non-relativistic approximation valid for $M\mu\ll1$ (see App.~\ref{app:non-rel}).
In practice, accurate results required for continuous GW data analysis are typically obtained numerically, for instance using Leaver's method~\cite{doi:10.1063/1.527130,Dolan:2007mj}.
Furthermore, corrections to the frequencies from various effects are also important and will be the main focus of the discussion below.

\section{Relativistic perturbation theory}\label{sec:perturbation}
In this section, we present a formulation of the frequency shift within relativistic perturbation theory.
In particular, we consider perturbations arising from self-interactions and self-gravity.

\subsection{Formulation}
\subsubsection{Bilinear product}
As the nonperturbative background, we begin with the Lagrangian density for the scalar field,
\begin{align}
    {\cal L}=-\frac{1}{2}\sqrt{-g}\left(g^{\mu\nu}\nabla_\mu\phi\nabla_\nu\phi+\mu^2\phi^2\right)~.
\end{align}
The equation of motion is given by Eq.~\eqref{eq:KGeq}, which we write as ${\cal O}\phi=0$.
In what follows, we work with the constant-$t$ hypersurface in Boyer--Lindquist coordinates.
Here, following Ref.~\cite{Cannizzaro:2023jle}, we define a bilinear product (related to the Klein--Gordon symplectic form) for the complex mode functions satisfying the equation of motion as
\begin{align}
    \langle\langle\Phi_i,\Phi_j\rangle\rangle=\int_{\Sigma_t}d\Sigma\  n^\mu\left({\cal J}\Phi_i\nabla_\mu\Phi_j-\Phi_j\nabla_\mu{\cal J}\Phi_i\right)~,
\end{align}
where $d\Sigma=d^3x\sqrt{-g}$ and $n_\mu dx^\mu=-dt$.
Here, we have introduced the symmetry operator ${\cal J}$ associated with the time-translation and axial symmetry in Kerr spacetime.
We define ${\cal J}$ to act on $\Phi_i$ as
\begin{align}
    {\cal J}\Phi_i=e^{+i(\omega_i t-m_i\varphi)}R_i^\ast(r)S_i^\ast(\theta)~.
\end{align}
Compared to Ref.~\cite{Cannizzaro:2023jle}, in addition to the transformation $(t,\varphi)\to(-t,-\varphi)$, we also take the complex conjugation of functions of $r$ and $\theta$.
Since ${\cal O}{\cal J}={\cal J}{\cal O}$, ${\cal J}\Phi_i$ is also the solution of the equation of motion.

This product is clearly a complex bilinear form.
Also, as long as $\Phi_i$ and $\Phi_j$ are solutions of the equation of motion and satisfy suitable decay condition, it does not depend on $\Sigma_t$, {\it i.e.}, it is conserved.
Furthermore, by applying the symmetry operator, the product satisfies the following useful properties:
\begin{enumerate}[label=(\roman*)]
    \item $\langle\langle\Phi_i,\Phi_j\rangle\rangle=\langle\langle\Phi_j,\Phi_i\rangle\rangle$,
    \item $\langle\langle{\cal L}_t\Phi_i,\Phi_j\rangle\rangle=\langle\langle\Phi_i,{\cal L}_t\Phi_j\rangle\rangle$ for $t^\mu$ the time translation Killing vector.
\end{enumerate}
These properties can be shown as follows~\cite{Green:2022htq}:
\begin{enumerate} [label=(\roman*)]
    \item Using ${\cal J}^2=1$ and the fact that ${\cal J}$ reverses the orientation of $\Sigma_t$,
    \begin{align}
        \langle\langle\Phi_i,\Phi_j\rangle\rangle&=\int _{\Sigma_t}d\Sigma\ n^\mu{\cal J}\left(\Phi_i\nabla_\mu({\cal J}\Phi_j)-{\cal J}\Phi_j\nabla_\mu\Phi_i\right) \notag \\
        &=-\int _{\Sigma_t}d\Sigma\ n^\mu\left(\Phi_i\nabla_\mu({\cal J}\Phi_j)-{\cal J}\Phi_j\nabla_\mu\Phi_i\right) \notag \\
        &=\langle\langle\Phi_j,\Phi_i\rangle\rangle~.
    \end{align}
    \item Using the conservation of the bilinear form and ${\cal L}_t{\cal J}=-{\cal J}{\cal L}_t$,
    \begin{align}
        0&={\cal L}_t\langle\langle\Phi_i,\Phi_j\rangle\rangle \notag \\
        &=\int _{\Sigma_t}d\Sigma\ n^\mu\left[-{\cal J}({\cal L}_t\Phi_i)\nabla_\mu\Phi_j+{\cal J}\Phi_i\nabla_\mu{\cal L}_t\Phi_j\right. \notag \\
        &\phantom{=\int _{\Sigma_t}d\Sigma\ n^\mu\left[\right.}\left.-{\cal L}_t\Phi_j\nabla_\mu{\cal J}\Phi_i+\Phi_j\nabla_\mu{\cal J}({\cal L}_t\Phi_i)\right] \notag \\
        &=-\langle\langle{\cal L}_t\Phi_i,\Phi_j\rangle\rangle+\langle\langle\Phi_i,{\cal L}_t\Phi_j\rangle\rangle~.
    \end{align}
\end{enumerate}

The properties derived above rely on the assumption that the radial part of the mode functions has compact support.
In Boyer--Lindquist coordinates, quasinormal modes diverge both at spatial infinity and at the horizon, while quasibound states with $\omega_I<0$ (the non-superradiant modes) diverge at the horizon.
Therefore, they cannot be directly applied to such modes.
In these cases, an appropriate extension is required, for example through a regularization in which the radial integration is deformed onto a complex contour.
By contrast, superradiant quasibound states have compact support and thus the bilinear form applies without modification.
Since the present work is concerned exclusively with these modes, we directly employ the bilinear form in its original form.

\subsubsection{Perturbation theory}
For the development of perturbation theory, it is convenient to move to the Hamiltonian formulation.
Here, we introduce the canonical momentum conjugate to the scalar field $\phi$,
\begin{align}
    \pi&=\frac{\partial{\cal L}}{\partial \dot{\phi}} \notag \\
    &=\sqrt{-g}\ n^\mu\nabla_\mu \phi~,
\end{align}
the Hamiltonian density is given by
\begin{align}
    {\cal H}&=\pi\dot{\phi}-{\cal L} \notag \\
    &=-\frac{1}{2\sqrt{-g}g^{tt}}\pi^2-\frac{g^{ti}}{g^{tt}}\pi\partial_i\phi \notag \\
    &\phantom{=}+\frac{1}{2}\sqrt{-g}\left(A^{ij}\partial_i\phi\partial_j\phi+\mu^2\phi^2\right)~.
\end{align}
where $A^{ij}=g^{ij}-g^{ti}g^{tj}/g^{tt}$.
Then, the equation of motion for the scalar field can be written as Hamilton's equation,
\begin{align}
    \dot{\phi}=\frac{\partial{\cal H}}{\partial \pi}~,\quad  {\rm and} \quad \dot{\pi}=-\frac{\partial {\cal H}}{\partial \phi}~.
\end{align}
Now, introducing phase-space vector
\begin{align}
    F=\begin{pmatrix}
        \phi \\ \pi
    \end{pmatrix}~,
\end{align}
these equations can be rewritten as
\begin{align}\label{eq:EoM_phase}
    {\cal L}_t F=HF~.
\end{align}
Here, $H$ can be interpreted as the effective Hamiltonian operator, whose explicit form is given by
\begin{align}\label{eq:Heff}
    H=\begin{pmatrix}
        -\frac{g^{ti}}{g^{tt}}\partial_i & \ -\frac{1}{\sqrt{-g}g^{tt}}\\
        \partial_i\left(\sqrt{-g}A^{ij}\partial_j\ \cdot\right)-\sqrt{-g}\mu^2 & \ -\partial_i\left(\frac{g^{ti}}{g^{tt}}\ \cdot\right)
    \end{pmatrix}~.
\end{align}
One can easily confirm that Eq.~\eqref{eq:EoM_phase} is equivalent to the Klein-Gordon equation~\eqref{eq:KGeq}.
This effective Hamiltonian operator is also equivalent to Eq.~(18) in Ref.~\cite{Green:2022htq}, where the same first-order system is expressed in the ADM form
\footnote{The one component form is also discussed in the Supplemental Material of Ref.~\cite{Green:2022htq}.}. Here, we use an explicit coordinate expression in terms of the metric components, which is convenient for deriving $\delta H$ induced by self-interaction and self-gravity effects in a unified notation.

The bilinear product in phase-space is defined as
\begin{align}\label{eq:product_phase}
    \langle\langle F_i,F_j\rangle\rangle=\int_{\Sigma_t}d^3x\left[({\cal J}\Phi_i)\Pi_j+({\cal J}\Pi_i)\Phi_j\right]~,
\end{align}
where $F_i=(\Phi_i,\Pi_i)^{\rm T}$ and $\Pi_i$ is the canonical momentum conjugate to $\Phi_i$.
Similarly, one can easily show that the product~\eqref{eq:product_phase} is equivalent to the one in the field space, {\it i.e.}, $\langle\langle F_i,F_j\rangle\rangle=\langle\langle \Phi_i,\Phi_j\rangle\rangle$.
Accordingly, it shares the same properties. In particular, using Eq.~\eqref{eq:EoM_phase}, the product in the phase-space satisfies
\begin{enumerate}[label=(\roman*)]
    \item $\langle\langle F_i,F_j\rangle\rangle=\langle\langle F_j,F_i\rangle\rangle$,
    \item $\langle\langle HF_i,F_j\rangle\rangle=\langle\langle F_i,HF_j\rangle\rangle$.
\end{enumerate}

Next, we consider perturbation theory for a perturbed Hamiltonian of the form $H\to H+\delta H$.
Since the equation of motion~\eqref{eq:EoM_phase} has been reduced to a first order differential equation in time, the discussion can proceed in close analogy with quantum mechanics.
If $\delta H$ is assumed to be approximately time independent, the perturbed mode vector can be expressed as $F_i\to F_i+\delta F_i$, which satisfy
\begin{align}
    (H+\delta H)(F_i+\delta F_i)=-i(\omega_i+\delta\omega_i)(F_i+\delta F_i)~,
\end{align}
where $\delta \omega_i$ is the frequency shift due to the perturbation $\delta H$.
Thus, at first order, we have
\begin{align}
    (H+i\omega_i)\delta F_i+\delta H F_i=-i\delta\omega_i F_i~.
\end{align}
We now take the inner product of both sides with $F_i$. Noting that
\begin{align}
    \langle\langle F_i,H\delta F_i\rangle\rangle=\langle\langle HF_i,\delta F_i\rangle\rangle=-i\omega_i\langle\langle F_i,\delta F_i\rangle\rangle~,
\end{align}
we obtain the frequency shift 
\begin{align}\label{eq:shift}
    \delta\omega_i=\frac{i\langle\langle F_i,\delta H F_i\rangle\rangle}{\langle\langle F_i,F_i\rangle\rangle}~.
\end{align}

In the following subsections, we consider specific forms of $\delta H$ and present explicit expressions for the associated frequency shifts.
In particular, since the cloud mass evolves much more slowly than the oscillation timescale, the above discussion remains applicable, and its variation can be treated adiabatically.
We also summarize the corresponding formulas in the non-relativistic approximation in App.~\ref{app:non-rel}, where it can be verified that the relativistic results agree with them in the appropriate limit.

\subsection{Self-interaction}
We first consider the perturbation induced by the axion self-interaction.
From Eq.~\eqref{eq:action_SI}, the perturbation of the Hamiltonian density is given by
\begin{align}
    \delta{\cal H}=-\sqrt{-g}\frac{1}{4!}\mu^2F_a^2\left(\frac{\phi}{F_a}\right)^4~.
\end{align}
For the effect on each mode, contributions from rapidly oscillating terms can be neglected in the time-averaged dynamics.
Thus, in the equation of motion for $\Phi_i$, the terms that contribute to the frequency shift arise from
\begin{align}
    \phi^3\supset\frac{1}{2\sqrt{2}}\left(3|\Phi_i|^2\Phi_i+6|\Phi_j|^2\Phi_i+\cdots\right)~.
\end{align}
Therefore the perturbation of the effective Hamiltonian sourced by the $j$-mode is given by
\begin{align}
    \delta H_j^{\rm SI}=\begin{pmatrix}
        0 & 0 \\
        \sqrt{-g}\frac{1}{2(1+\delta_{ij})}\frac{\mu^2}{F_a^2}|\Phi_j|^2 & 0
    \end{pmatrix}~.
\end{align}
Here, substituting Eq.~\eqref{eq:mode}, the norm can be expressed more explicitly as
\begin{align}
    &\langle\langle F_i,F_i\rangle\rangle \notag \\
    &=2iM_{c,i}\int d^3x\sqrt{-g}(g^{tt}\omega_i-g^{t\varphi}m_i)|R_i(r)|^2|S_i(\theta)|^2~.
\end{align}
Here and in what follows, we neglect the imaginary part of $\omega_i$, since it is sufficiently small.
On the other hand, the product appearing in the numerator of Eq.~\eqref{eq:shift} is given by
\begin{align}
    &\langle\langle F_i,\delta H_j^{\rm SI}F_i\rangle\rangle \notag \\
    &=\frac{1}{2(1+\delta_{ij})}\frac{\mu^2}{F_a^2}M_{c,i}M_{c,j} \notag \\
    &\phantom{=}\times\int d^3x\sqrt{-g}|R_i(r)|^2|R_j(r)|^2|S_i(\theta)|^2|S_j(\theta)|^2~.
\end{align}
Substituting these expressions, we obtain the frequency shift due to the self-interaction,
\begin{align}\label{eq:shift_SI}
    \delta\omega_{ij}^{\rm SI}=\frac{i\langle\langle F_i,\delta H_j^{\rm SI} F_i\rangle\rangle}{\langle\langle F_i,F_i\rangle\rangle}~.
\end{align}

\subsection{Self-gravity}
Next, we consider the perturbation induced by the self-gravity of the axion cloud.
In this case, the metric is perturbed as $g_{\mu\nu}\to g_{\mu\nu}+h_{\mu\nu}$.
Therefore, from Eq.~\eqref{eq:Heff}, the perturbed effective Hamiltonian is given by
\begin{widetext}
\begin{align}
    \delta H=\begin{pmatrix}
        \left(\frac{h^{ti}}{g^{tt}}-\frac{g^{ti}}{(g^{tt})^2}h^{tt}\right)\partial_i & \ \frac{1}{\sqrt{-g}g^{tt}}\left(\frac{h}{2}-\frac{h^{tt}}{g^{tt}}\right) \\
        \partial_i\left[\sqrt{-g}\left(\delta A^{ij}+\frac{h}{2}A^{ij}\right)\partial_j \ \cdot\right]-\frac{1}{2}\sqrt{-g}h\mu^2 & \ \partial_i\left[\left(\frac{h^{ti}}{g^{tt}}-\frac{g^{ti}}{(g^{tt})^2}h^{tt}\right)\ \cdot\right]
    \end{pmatrix}~,
\end{align}
\end{widetext}
where $\delta A^{ij}=-h^{ij}+(h^{ti}g^{tj}+h^{tj}g^{ti})/g^{tt}-h^{tt}g^{ti}g^{tj}/(g^{tt})^2$ and $h=g^{\mu\nu}h_{\mu\nu}$.
We note that, for the variation $\delta_\xi H$ induced by the gauge transformation $h_{\mu\nu}\to h_{\mu\nu}+2\nabla_{(\mu}\xi_{\nu)}$, $\langle\langle F_i,\delta_\xi HF_i\rangle\rangle=0$.
In other words, the frequency shift is gauge invariant.
This is shown in App.~\ref{app:gauge_inv}.

The metric perturbation $h_{\mu\nu}$ is obtained as a solution to the linearized Einstein equation sourced by Eq.~\eqref{eq:energy_momentum_tensor}.
In principle, it is possible to perform the calculation on a Kerr background by means of the metric reconstruction technique~\cite{Hollands:2024iqp,Wardell:2024yoi}.
However, since it is technically challenging, we instead follow Ref.~\cite{Cannizzaro:2023jle} and adopt the semi-Newtonian approximation here.
In other words, the metric
\begin{align}
    ds^2=-(1+2U)dt^2+(1-2U)(dr^2+r^2d\Omega^2)~,
\end{align}
is employed only for $\delta H$.
Here, $U$ is the gravitational potential sourced by the self-gravity and satisfies
\begin{align}\label{eq:Poisson}
    \nabla^2U=4\pi T_{tt}~,
\end{align}
where $\nabla^2$ is the Laplacian in flat space.

For the source term, we denote the contribution from the $i$-mode by $T_{tt}[\Phi_i]$, which is given by
\begin{align}
    T_{tt}[\Phi_i]=\omega_i^2|\Phi_i|^2-\frac{1}{2}g_{tt}\left(g^{\mu\nu}\partial_\mu\Phi_i\partial_\nu\Phi_i^\ast+\mu^2|\Phi_i|^2\right)~.
\end{align}
Solving Eq.~\eqref{eq:Poisson}, the corresponding gravitational potential is obtained as
\begin{align}\label{eq:gravitational_potential}
    U_i=&-\sum_l\frac{4\pi}{2l+1}Y_{l0}(\Omega) \notag \\
    &\phantom{-\sum_l}\times\int d^3x'T_{tt}[\Phi_i]Y_{l0}^\ast(\Omega') \notag \\
    &\phantom{-\sum_l\times\int}\times\left(\frac{r^{\prime l}}{r^{l+1}}\Theta(r-r')+\frac{r^l}{r^{\prime l+1}}\Theta(r'-r)\right)~,
\end{align}
where $Y_{lm}$ are the spherical harmonics, and $\Theta$ is the Heaviside step function.
Since $T_{tt}[\Phi_i]$ contains only the $m=0$ component, only the $m=0$ component of the spherical harmonics survives.

In the semi-Newtonian approximation, the perturbed effective Hamiltonian induced by the $j$-mode self-gravity reduces to
\begin{align}
    \delta H_j^{\rm SG}=\begin{pmatrix}
        0 & \frac{1}{\sqrt{-g}}4U_j \\
        2\sqrt{-g}\mu^2 U_j & 0
    \end{pmatrix}~.
\end{align}
Compared with Ref.~\cite{Cannizzaro:2023jle}, where $\delta H=U H$ was derived, the present derivation also includes the variation of the spatial components of the metric.
In addition, in this work we treat $T_{tt}$ relativistically and include contributions to $U$ from higher-$l$ multipoles in the actual calculation.

Then, the product appearing in the numerator of Eq.~\eqref{eq:shift} is given by
\begin{align}
    \langle\langle F_i,\delta H_j^{\rm SG}F_i\rangle\rangle=\int d^3x\sqrt{-g}&\left(2\mu^2-4(g^{tt}\omega_i-g^{t\varphi}m_i)^2\right) \notag \\
    &\times U_j|R_i(r)|^2|S_i(\theta)|^2~.
\end{align}
Substituting these expressions, we finally obtain the frequency shift due to the self-gravity,
\begin{align}\label{eq:shift_SG}
    \delta\omega_{ij}^{\rm SG}=\frac{i\langle\langle F_i,\delta H_j^{\rm SG} F_i\rangle\rangle}{\langle\langle F_i,F_i\rangle\rangle}~.
\end{align}

\section{Numerical results}\label{sec:numerical}
In this section, we present numerical results for the frequency shifts formulated above.
In particular, we compare them with results obtained by other methods and discuss their impact on the GW frequencies.

\subsection{Comparison}

\begin{figure}[t]
\centering
\includegraphics[width=0.9\linewidth]{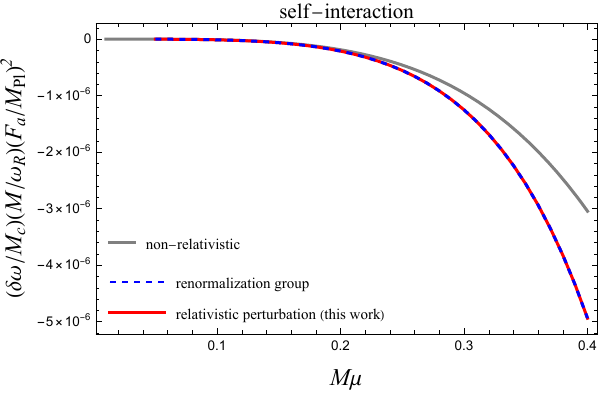}
\caption{Comparison of the frequency shift induced by self-interaction. The mode of interest is $(n,l,m)=(2,1,1)$, and the BH spin is set to $a/M=a_{\rm cr}/M+10^{-2}$. The gray solid line denotes the non-relativistic approximation, the blue dashed line the renormalization group result~\cite{Omiya:2020vji}, and the red line the relativistic perturbation result.}
\label{fig:comparison_SI}
\end{figure}

First, we discuss the results for the shift induced by self-interaction.
Fig.~\ref{fig:comparison_SI} shows the shift of the $(n,l,m)=(2,1,1)$ mode induced by the self-interaction of the $(2,1,1)$ mode itself.
In this plot, the BH spin is set to a value slightly above the critical spin $a_{\rm cr}/M=4m(M\mu)/(m^2+4(M\mu)^2)$
\footnote{Strictly speaking, $\mu$ should be replaced by $\omega_R$. However, since this is merely a choice of parameter setting, we adopt this expression here.}
.
As can be derived from the analytic expression (see App.~\ref{app:non-rel}), the result obtained from relativistic perturbation theory agrees well with the non-relativistic approximation in the small-$M\mu$ regime.

We also compared the results of this work with those obtained from the analysis based on the renormalization group method developed in Ref.~\cite{Omiya:2020vji}.
In that reference, a formula valid up to second order in the cloud mass (or quartic order in the amplitude) was derived (Eq.~(66)).
Here, however, we focus only on the first order contribution in order to compare with the formulation presented in this paper.
As a result, as shown in Fig.~\ref{fig:comparison_SI}, these results are in almost perfect agreement.

The advantage of the renormalization group method is that a formulation has been established up to second order in the cloud mass. Also, full nonlinear effects can be obtained using the method developed in Ref.~\cite{Omiya:2022mwv}, which solves the adiabatic evolution including changes in the cloud configurations.
On the other hand, the perturbative approach presented here is simpler both in its derivation and practical implementation.
Furthermore, it has the advantage that it can be directly applied to couplings with other modes, which are not incorporated in the former approach.

\begin{figure}[t]
\centering
\includegraphics[width=0.9\linewidth]{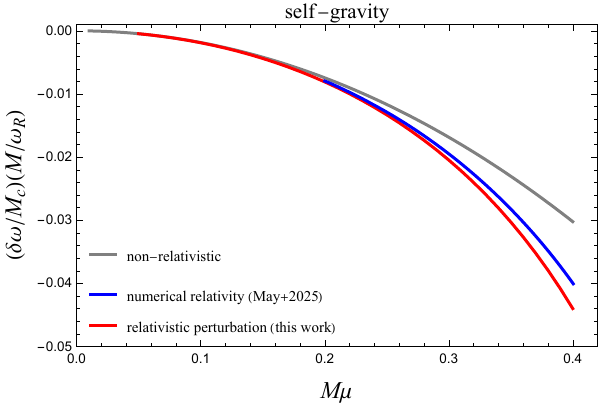}
\caption{Comparison of the frequency shift induced by self-gravity. The mode of interest is $(n,l,m)=(2,1,1)$, and the BH spin is set to $a/M=a_{\rm cr}/M+10^{-2}$. The gray solid line denotes the non-relativistic approximation, the blue solid line the result extracted from numerical relativity simulation~\cite{May:2024npn}, and the red line the relativistic perturbation result.}
\label{fig:comparison_SG}
\end{figure}

Next, we discuss the results for the shift induced by self-gravity.
Fig.~\ref{fig:comparison_SG} shows the results for the same choice of parameters and mode configurations as in the self-interaction case.
Similarly, we confirm the results agree well with the non-relativistic approximation in the small-$M\mu$ regime.

In addition, we compared our results with the numerical relativity simulation results reported in Ref.~\cite{May:2024npn}.
From the publicly available \texttt{SuperRad}~\cite{Siemonsen:2022yyf}, we extracted the coefficient of the frequency shift at first order in the cloud mass, and show it together with the other results in Fig.~\ref{fig:comparison_SG}.
As in Ref.~\cite{Cannizzaro:2023jle}, the relativistic perturbation result is found to be closer to the simulation result than the non-relativistic approximation.
However, even within our formulation, there remains a relative error of $10\%$ at $M\mu=0.4$.
This discrepancy is likely due to the Newtonian approximation used for the gravitational perturbation.
Further improvement is expected by employing BH perturbation theory for the metric perturbation, but we leave this for future work.

\subsection{GW frequency}

\begin{figure}[t]
\centering
\includegraphics[width=0.9\linewidth]{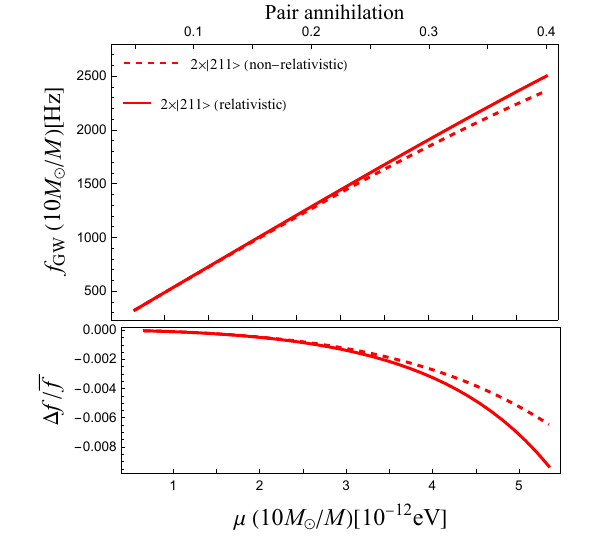}
\includegraphics[width=0.9\linewidth]{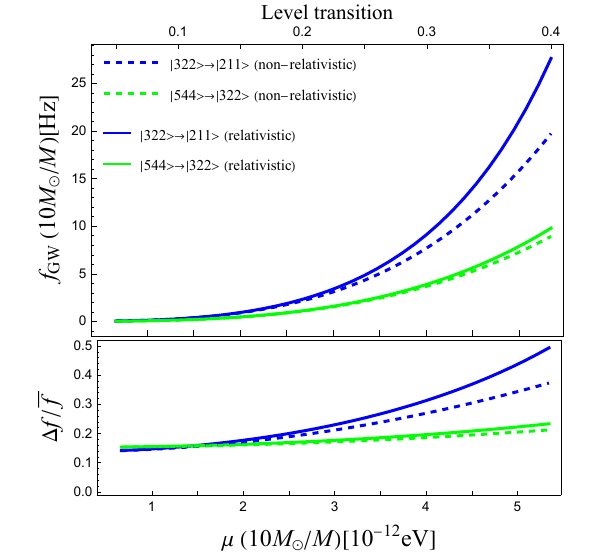}
\caption{GW frequencies including frequency shifts, together with their relative differences from the cases without the shift. The upper panel shows GWs from pair annihilation, while the lower panel shows those from the level transitions. The dashed lines represent the non-relativistic approximation, and the solid line represents the relativistic perturbation results. The masses of the clouds involved in the emission processes are all set to $M_c=0.1M$, the decay constant is set to $F_a=10^{-2}M_{\rm Pl}$, and the BH spin is set to $a/M=0.99$. The labels on the top axis indicate the values of $M\mu$.}
\label{fig:GWfrequency}
\end{figure}

Finally, we present explicit examples of the GWs actually emitted from axion clouds.
First, we consider GWs from the pair annihilation process of the $(2,1,1)$ mode.
Assuming that only this mode is excited, the frequency is given by
\begin{align}
    \omega_{1}=\bar{\omega}_{1}+\delta\omega_{11}^{\rm SI}+\delta\omega_{11}^{\rm SG}~,
\end{align}
where $\delta\omega_{ij}$ represents the shift in the eigenfrequency of the $(i+1,i,i)$ mode induced by the $(j+1,j,j)$ mode.
Here and in what follows, quantities with an overbar denote unperturbed ones.
In this process, the GWs are produced by two axions, and hence the GW frequency is given by $f_{\rm GW}=\omega_1/\pi$.

On the other hand, as GWs from the level transition, we consider the radiation associated with the transition $(3,2,2)\to(2,1,1)$ and $(5,4,4)\to(3,2,2)$.
In particular, the latter can provide the strongest signal~\cite{Omiya:2024xlz}.
Assuming that only the two modes involved in the transition are excited, the frequency of each mode is given by
\begin{align}
    \omega_i=\bar{\omega}_i+\delta\omega_{ii}^{\rm SI}+\delta\omega_{ij}^{\rm SI}+\delta\omega_{ii}^{\rm SG}+\delta\omega_{ij}^{\rm SG}~.
\end{align}
From this, the GW frequency is expressed as $f_{\rm GW}=(\omega_j-\omega_i)/2\pi$.

Fig.~\ref{fig:GWfrequency} shows the GW frequencies for each axion mass arising from the processes described above.
In these plots, the masses of all relevant clouds are set to $M_c=0.1 M$, the decay constant to $F_a=10^{-2}M_{\rm Pl}$ (roughly the GUT scale), and the BH spin to $a/M=0.99$.
In addition, the relative difference from the unperturbed case, $\Delta f/\bar{f}$, is shown below each frequency, where $\Delta f=f_{\rm GW}-\bar{f}$.

As can be seen, the relativistic results can be significantly different from the non-relativistic ones when $M\mu$ is not small.
However, this is only one example of a parameter set, and we note that the behavior strongly depends on $(M\mu,M_c,F_a)$.
In particular, the effect on the frequency shift is pronounced for GWs from level transitions.
This is because, even if the shift in each mode frequency is small, the GW frequency is determined by their difference.

Finally, let us comment on the relative importance of self-interaction and self-gravity effects. As shown in App.~\ref{app:non-rel}, within the non-relativistic approximation, the ratio of the two corrections is estimated as
\begin{align}
    \frac{\delta\omega^{\rm SI}}{\delta\omega^{\rm SG}}\sim\left(\frac{M\mu}{M_{\rm Pl}^2}\right)^2\left(\frac{M_{\rm Pl}}{F_a}\right)^2~,
\end{align}
where we have restored the Planck mass $M_{\rm Pl}=1/\sqrt{G}$.
Thus, the self-interaction contribution is expected to dominate roughly when $F_a/M_{\rm Pl}\lesssim M\mu/M_{\rm Pl}^2$. This estimate shows that stronger self-interactions, corresponding to smaller $F_a$, can make the frequency shift from self-interactions more important than that from self-gravity. In practice, however, the numerical prefactors are mode dependent and can differ significantly between the two effects. Therefore, which effect gives the dominant correction depends sensitively on the parameters and on the gravitational-wave emission process under consideration.

For example, for the parameter choice used in Fig.~\ref{fig:GWfrequency}, the self-interaction contribution becomes dominant for $M\mu\gtrsim0.38$ in the (2,1,1) pair annihilation process, and for $M\mu\gtrsim0.31$ in the (3,2,2) $\to$ (2,1,1) level transition. For the (5,4,4) $\to$ (3,2,2) level transition, self-gravity remains dominant over the range $M\mu\leq0.4$ considered here, but the self-interaction contribution is expected to dominate at larger values of $M\mu$.

\section{Conclusions}\label{sec:conc}
In this paper, we investigated perturbative shifts in the frequencies of axion clouds within a relativistic framework.
These shifts are directly related to the frequencies of GWs emitted from the axion cloud.
We first introduced a bilinear product with desirable properties and formulated perturbation theory by moving to a Hamiltonian framework.
In particular, we considered axion self-interaction and self-gravity as perturbations, and derived explicit expressions for the resulting frequency shifts.
This makes it possible to compute frequency shifts in a simple manner in situations where multiple modes are excited, which is important when considering various GW emission processes.

We also carried out several explicit calculations of the frequency shifts formulated in this work.
As can be derived from the analytic expressions, the results are confirmed to agree well with the non-relativistic approximation in the small-$M\mu$ regime.
We further compared our results with previous results obtained by other methods.
For self-interaction, we compared with the renormalization group method developed in Ref.~\cite{Omiya:2020vji}.
As a result, the first-order contribution in the cloud mass was found to be in almost perfect agreement.
For self-gravity, we compared with the numerical relativity simulation results presented in Ref.~\cite{May:2024npn}.
In this case, a relative difference of $10\%$ remained at $M\mu=0.4$.
As in Ref.~\cite{Cannizzaro:2023jle}, this discrepancy is likely due to the Newtonian treatment of the gravitational perturbation.
Possible future improvements include extending the analysis to second order perturbation theory in order to incorporate higher-order effects of the cloud mass, and treating the gravitational perturbation within BH perturbation theory.

We also presented the GW frequencies from pair annihilation and level transitions after taking into account realistic emission processes.
In particular, for the latter case, the effect of the frequency shift is significant since the GW frequency is given by the difference between the mode frequencies of the axion cloud.
In practice, the most relevant targets for current and future ground-based GW detectors, as well as space-based detectors, lie in the relativistic regime of large $M\mu$.
In this regime, relativistically computed frequency shifts become important.
For future observations with improved sensitivity, it is expected that the constraints will be extended to parameter regions where the GW signal is weaker.
For instance, in regions with strong self-interaction, dissipation reduces the cloud mass, which results in a smaller GW amplitude.
In such regions, the contribution of self-interaction to the frequency shift can become dominant.
A concrete estimate of detectability, however, requires fixing the search strategy and lies beyond the scope of this work.
Detailed analyses using current data, as well as studies of detectability with future observations, will be pursued in future work.
Finally, we mention several possible directions for theoretical extensions, including applications to vector clouds, tidal interactions and corrections to the quasinormal mode spectrum.


\acknowledgments{
I would like to thank Taillte May and Hidetoshi Omiya for helpful correspondence.
This work is supported by JSPS (Japan Society for the Promotion of Science) KAKENHI Grant Nos.~JP25KJ0067, and JP25K17397.
}


\appendix

\section{Non-relativistic approximation}\label{app:non-rel}
In this appendix, we summarize the formulas for the frequency shifts in the non-relativistic approximation.
In the regime $M\mu\ll1$ and $a/M\ll1$, we adopt the ansatz
\begin{align}
    \phi=\frac{1}{\sqrt{2\mu}}\left(e^{-i\mu t}\psi+e^{i\mu t}\psi\right)~,
\end{align}
where $\psi(t,x^i)$ is a complex function whose timescale of variation is much longer than $\mu^{-1}$.
Ignoring the second time derivative of $\psi$, the equation of motion becomes
\begin{align}
    \left(i\partial_t+\frac{1}{2\mu}\nabla^2+\frac{M\mu}{r}\right)\psi=\delta V\psi~.
\end{align}
Here, $\delta V$ represents the perturbation that sources the frequency shift.
The unperturbed eigenmodes can be written as
\begin{align}
    \psi_i=\sqrt{\frac{M_{c,i}}{\mu}}e^{-i(\omega_i-\mu)t}R_i(r)Y_i(\Omega)~,
\end{align}
where $R_i(r)$ are the wavefunctions of a hydrogen atom.
In this approximation, the bilinear form reduces to the standard $L^2$ inner product.
Thus, as in the main text, the frequency shift is given by
\begin{align}
    \delta\omega_i=\frac{\int d^3\bm{x} \psi^\ast_i \delta V\psi_i}{\int d^3\bm{x}|\psi_i|^2}~,
\end{align}
where $d^3\bm{x}=r^2\sin\theta dr d\theta d\varphi$.

For the perturbation sourced by the $j$-mode, the self-interaction contribution is given by
\begin{align}
    \delta V_j^{\rm SI}=-\frac{1}{4(1+\delta_{ij})}\frac{1}{F_a^2}|\psi_j|^2~,
\end{align}
while the self-gravity contribution is given by
\begin{align}
    \delta V_j^{\rm SG}=\mu U_j~.
\end{align}
Here, $U_j$ is given by Eq.~\eqref{eq:gravitational_potential}, where $T_{tt}\simeq \mu|\psi_j|^2$ in the present approximation.
Consequently, with the present normalization, $\int d^3\bm{x}|\psi_i|^2=M_{c,i}/\mu$.

Therefore, we obtain the following expressions for the frequency shifts:
\begin{align}
    \delta\omega_{ij}^{\rm SI}=&-\frac{1}{4(1+\delta_{ij})}\frac{1}{F_a^2}\frac{M_{c,j}}{\mu} \notag \\
    &\times\int d^3\bm{x}|R_i(r)|^2|R_j(r)|^2|Y_i(\Omega)|^2|Y_j(\Omega)|^2~,
\end{align}
\begin{align}
    \delta\omega_{ij}^{\rm SG}=&-\mu M_{c,j}\sum_l\frac{4\pi}{2l+1} \notag \\
    &\times\int d^3\bm{x}d^3\bm{x}'|R_i(r)|^2|R_j(r')|^2 \notag \\
    &\times\left(\frac{r^{\prime l}}{r^{l+1}}\Theta(r-r')+\frac{r^l}{r^{\prime l+1}}\Theta(r'-r)\right) \notag \\
    &\times Y_{l0}(\Omega)|Y_i(\Omega)|^2Y_{l0}^\ast(\Omega')|Y_j(\Omega')|^2~.
\end{align}
In particular, in the non-relativistic approximation, the configuration of the mode functions is independent of the eigenfrequency, and therefore the parametric dependence can be factored out of the overlap integral.
Thus, the frequency shifts can be written as
\begin{align}
    &\delta\omega_{ij}^{\rm SI}=-\kappa_{ij}^{\rm SI}\mu\left(\frac{M\mu}{M_{\rm Pl}^2}\right)^4\left(\frac{M_{\rm Pl}}{F_a}\right)^2\frac{M_{c,j}}{M}~, \\
    &\delta\omega_{ij}^{\rm SG}=-\kappa_{ij}^{\rm SG}\mu\left(\frac{M\mu}{M_{\rm Pl}^2}\right)^2\frac{M_{c,j}}{M}~,
\end{align}
where we have restored the explicit dependence on the Planck mass $M_{\rm Pl}$.
The coefficients of the frequency shifts relevant to the main text are summarized in Table.~\ref{tab:coeff}.

\begin{table}[t]
\caption{Coefficients of the frequency shifts. $\kappa_{ij}$ represents the coefficient associated with the shift in the eigenfrequency of the $(n,l,m)=(i+1,i,i)$ mode induced by the source generated by the $(n,l,m)=(j+1,j,j)$ mode.}
\label{tab:coeff}
\begin{ruledtabular}
\begin{tabular}{cccc}
 & Self-interaction &  & Self-gravity \\
\colrule
$\kappa^{\rm SI}_{11}$ & $1.2\times10^{-4}$ &
$\kappa^{\rm SG}_{11}$ & $1.9\times10^{-1}$ \\

$\kappa^{\rm SI}_{12}$ & $3.5\times10^{-5}$ &
$\kappa^{\rm SG}_{12}$ & $1.1\times10^{-1}$ \\

$\kappa^{\rm SI}_{22}$ & $1.4\times10^{-5}$ &
$\kappa^{\rm SG}_{22}$ & $9.0\times10^{-2}$ \\

$\kappa^{\rm SI}_{24}$ & $9.7\times10^{-7}$ &
$\kappa^{\rm SG}_{24}$ & $4.1\times10^{-2}$ \\

$\kappa^{\rm SI}_{44}$ & $1.1\times10^{-6}$ &
$\kappa^{\rm SG}_{44}$ & $3.6\times10^{-2}$ \\
\end{tabular}
\end{ruledtabular}
\end{table}

\section{Gauge invariance}\label{app:gauge_inv}
We now demonstrate the invariance of the frequency shift in Eq.~\eqref{eq:shift} under the gauge transformation of the metric perturbation.
Let the metric perturbation be generated by an infinitesimal vector field $\xi^\mu$ as
\begin{align}
    h_{\mu\nu}={\cal L}_\xi g_{\mu\nu}~.
\end{align}
Under the same infinitesimal diffeomorphism, the variation of the phase-space vector is given by
\begin{align}\label{eq:gauge_F}
    \delta_\xi F=-{\cal L}_\xi F~.
\end{align}
Since the Klein-Gordon equation is diffeomorphism invariant, we have
\begin{align}
    {\cal L}_t(F+\delta_\xi F)=(H+\delta_\xi H)(F+\delta_\xi F)~,
\end{align}
where $\delta_\xi H$ denotes the induced perturbation of the effective Hamiltonian.
Expanding this equation to first order, we obtain
\begin{align}\label{eq:gauge_first}
    {\cal L}_t\delta_\xi F=H\delta_\xi F+\delta_\xi H F~.
\end{align}
On the other hand, since the background spacetime is stationary, we may choose $\xi^\mu$ such that $[{\cal L}_t,{\cal L}_\xi]=0$ and
\begin{align}
    {\cal L}_t\delta_\xi F=-{\cal L}_t{\cal L}_\xi F=-{\cal L}_\xi{\cal L}_t F=-{\cal L}_\xi HF~.
\end{align}
Substituting this into Eq.~\eqref{eq:gauge_first}, we can write
\begin{align}
    \delta_\xi H=[H,{\cal L}_\xi]~.
\end{align}
Thus, 
\begin{align}
    \langle\langle F_i,\delta_\xi H F_i\rangle\rangle&=\langle\langle F_i,[H,{\cal L_\xi}] F_i\rangle\rangle \notag \\
    &=\langle\langle H F_i,{\cal L}_\xi F_i\rangle\rangle-\langle\langle F_i,{\cal L}_\xi HF_i\rangle\rangle \notag \\
    &=-i(\omega_i-\omega_i)\langle\langle F_i,{\cal L}_\xi F_i\rangle\rangle \notag \\
    &=0~.
\end{align}
Hence, the frequency shift is gauge invariant.


\bibliographystyle{mybibstyle}
\bibliography{ref}

\end{document}